\documentclass[epj,english,final,amsmath,amssymb]{svjour}
\usepackage{graphicx}
\input{epsf}
\begin{document}
\title{Magnetization of ballistic quantum dots \\ 
induced by a linear-polarized microwave field}

\author{A. D. Chepelianskii\inst{1}  and D. L. Shepelyansky\inst{2}}

\institute {Ecole Normale Sup\'erieure, 45, rue d'Ulm, 
75231 Paris Cedex 05, France
\and
Laboratoire de Physique Th\'eorique,
UMR 5152 du CNRS, Univ. P. Sabatier, 31062 Toulouse Cedex 4, France}

\titlerunning{Magnetization of quantum dots induced by a microwave field}
\authorrunning{A.D.Chepelianskii and D.L.Shepelyansky}

\date{Received:}
\abstract{On a basis of extensive analytical and numerical
studies we show that a linear-polarized microwave field
creates a stationary magnetization in mesoscopic ballistic quantum 
dots with two-dimensional electron gas being at a thermal equilibrium. 
The magnetization is proportional to a number of electrons 
in a dot and to a microwave power.
Microwave fields of moderate strength create  
in a one dot of few micron size a magnetization
which is by few orders of magnitude larger than
a magnetization produced by persistent currents.
The effect is  weakly dependent on temperature and
can be observed with existing experimental techniques.
The parallels between this effect and ratchets
in asymmetric nanostructures are also discussed.
}

\date{October 25, 2006}

\maketitle


\section{Introduction}
Since 1990, when magnetization of an ensemble of $10^7$ mesoscopic rings
had been detected experimentally
\cite{levy1990}, the problem of magnetization of quantum dots
of two-dimensional electron gas (2DEG)
or persistent currents attracted a great deal of attention
(see e.g. \cite{imry,gilles} and Refs. therein).
It is well known that a magnetic field
gives no magnetization in a classical system at thermal
equilibrium (see e.g. \cite{mayer}).
Thus, the persistent currents have a quantum origin
and are relatively weak corresponding to values
of one electron current of typical strength $3 \times {10}^{-3} e v_F/L_p$
\cite{levy1990} where $v_F$ is the Fermi velocity and $L_p$ is a perimeter of 
quantum dot. Due to that skillful experimental efforts 
had been required to detect persistent currents of strength
$I_0= e v_F/L_p$ in an isolated ballistic ring \cite{mailly}.

The effects of {\it ac}-field on magnetization of
mesoscopic dots have been discussed in \cite{kravtsov1993,kravtsov2005}.
It was shown that {\it ac}-driving induces persistent currents
which strength oscillates with magnetic flux. The amplitude of
the current is proportional to the intensity of microwave
field but still its amplitude is small compared to
one electron current $ e v_F/L_p$. 
Such currents have pure quantum origin and are
essentially given only by one electron on a quantum level
near the Fermi level. Experimental investigations
of magnetization induced by {\it ac}-driving have been
reported in \cite{bouchiat1,bouchiat2}. 
The amplitude of induced currents was in a qualitative 
agreement with the theoretical predictions \cite{kravtsov1993,kravtsov2005}.
However, the strength of currents in one ring was rather weak
(less than $nA$) and it was necessary to use an ensemble of ${10}^5$ rings
to detect induced magnetization.

In this work we show that a liner-polarized microwave field
generates strong orbital currents and magnetization
in ballistic quantum dots. The magnetization
(average non-zero momentum) of a dot is
proportional to the number of electrons inside the dot
and to the intensity of microwave field.
The sign of magnetization depends on orientation
of polarization in respect to symmetry axis of the dot.
It is assumed that without microwave field the dot is in
a thermal equilibrium characterized by the Fermi-Dirac
distribution with a temperature $T$. 
A microwave field drives  the system to
a new stationary state with non-zero stationary
magnetization to which contribute all electrons 
inside the dot. The steady state appears as 
a result of equilibrium between energy growth
induced by a microwave field and relaxation which
drives the system to thermal equilibrium.
In contrast to the persistent currents
discussed in \cite{imry,gilles,kravtsov2005} 
the dynamical magnetization effect 
discussed here has essentially classical origin
and hence it gives much stronger currents.
However, it disappears in the presence of disorder
when the mean free path  becomes smaller than
the size of the dot. 
Thus the electron dynamics should be ballistic inside
the dot which should have a stretched form
since the effect is absent inside a ring
and is weak inside a square shaped dot.
Also the effect is most strong when the
microwave frequency is comparable
with a frequency of electron oscillations inside the dot.
The above conditions were not fulfilled in the
experiments \cite{bouchiat1,bouchiat2}
and therefor the dynamical magnetization had not be seen there.

The physical origin of dynamical magnetization
can be seen already from a simple model
of two decoupled dissipative oscillators
for which a monochromatic driving 
leads to a certain degree of synchronization
with the driving phase  \cite{pikovsky}.
This phenomenological approach 
was proposed by Magarill and Chaplik
\cite{chaplik} who gave first estimates
for photoinduced magnetism in ballistic nanostructures.
Here we develop a more rigorous approach
based on the density matrix and semiclassical
calculations of dynamical magnetization.
We also extend our analysis to a generic 
case of enharmonic potential inside the dot
that leads to qualitative changes 
in the  magnetization dependence on microwave frequency.
We also trace certain parallels
between dynamical magnetization,
directed transport (ratchets)
induced by microwave fields in asymmetric
nanostructures \cite{ratchet1,ratchet2,alik,entin}
and the Landau damping \cite{landau,pitaevsky}.

The paper has the following structure:
in Section II we consider the case of a quantum dot with
harmonic potential, enharmonic potential is analyzed in
Section III, a quantum dot in a form of Bunimovich
stadium \cite{bunimovich} is considered in Section IV,
discussions and conclusions are given in Section V.  

\section{Quantum dots with a harmonic potential}

Electron dynamics inside a two-dimensional (2D) dot is described by
a Hamiltonian
\begin{equation}
\label{eq1}
H=(p_x^2+p_y^2)/2m + U(x,y) - x f_x \cos \omega t  - y f_y \cos \omega t 
\end{equation}
where $m$ is electron mass and $p_{x,y}$ and $x,y$
are conjugated momentum and coordinate. An external
force $f_{x,y}$ is created by a linear-polarized
microwave field with frequency $\omega$.
The polarization angle $\theta$ and the force amplitude $f$
are defined by relations $f_x=f \cos \theta$,
$f_y=f \sin \theta$.
In this Section we consider the case of a harmonic potential
$U(x,y) = m(\omega_x^2 x^2+\omega_y^2 y^2)/2 $
where $\omega_{x,y}$
are oscillation frequencies in $x,y$ directions
(generally non equal).

Let us consider first a phenomenological case
when an electron experiences an additional
friction force $\bf{ F} = - \gamma {\bf p}$
where $\gamma$ is a relaxation rate (this approach had been 
considered in \cite{chaplik} and we give it here only
for completeness). The dynamical equations of motion
in this case are linear and can be solved exactly that gives
at $t \gg 1/\gamma$:
\begin{equation}
\label{eq2}
    \begin{array}{ll}
  x(t) = \Re \frac{ e^{i \omega t} f_x/m }{\omega_x^2 - \omega^2 + i \gamma \omega} = \Re \; X(t) , \\
  y(t) = \Re \frac{ e^{i \omega t}f_y/m }{\omega_y^2 - \omega^2 + i \gamma \omega} = \Re \; Y(t) ,
    \end{array}
\end{equation}
where $\Re$ marks the real part. Then, the electron velocities are
\begin{equation}
\label{eq3}
    \begin{array}{ll}
  x(t) = \Re \frac{ i e^{i \omega t} \omega f_x/m }{\omega_x^2 - \omega^2 + i \gamma \omega} = \Re \; X(t) , \\
  y(t) = \Re \frac{ i e^{i \omega t} \omega f_y/m }{\omega_y^2 - \omega^2 + i \gamma \omega} = \Re \; Y(t) .
    \end{array}
\end{equation}
This gives the average momentum
\begin{eqnarray}
\nonumber
&& L  =  m \left \langle x(t) v_y(t) - y(t) v_x(t) \right \rangle  = \\
\label{eq4}
&& \Re \frac{- i \omega f_x f_y/m}{(\omega_x^2 - \omega^2 + i \gamma \omega)(\omega_y^2 - \omega^2 - i \gamma \omega)} \quad .
\end{eqnarray}
In  the limit of small $\gamma \ll \omega_{x,y}$
Eq.(\ref{eq4}) gives for the off-resonance case
\begin{equation}
\label{eq5}
L_{off} = \frac{\gamma \omega^2 (\omega_x^2 - \omega_y^2) f_x f_y/m}{(\omega_x^2 - \omega^2)^2(\omega_y^2 - \omega^2)^2} ,
\end{equation}
while at the resonance $\omega=\omega_x$
\begin{equation}
\label{eq6}
L_{res} = \frac{f_x f_y/m}{\gamma (\omega_x^2 - \omega_y^2)} .
\end{equation}
From a physical viewpoint an average momentum appears due to 
a phase shift between oscillator phases induced by dissipation
and an orbit takes an elliptic form with rotation in one direction.
In some sense, due to dissipation the two oscillators become synchronized by
external force \cite{pikovsky}.
As usual \cite{mayer}, an average orbital momentum $L$ for one electron 
gives a total magnetic moment $M=N L e/2mc$ where $N$
is a number of electrons in a quantum dot.
A pictorial view of spectral dependence of $M$ on $\omega$ at various ratios
$\omega_x/\omega_y$ is given in \cite{chaplik}. 

To extend the phenomenological approach described above
we should take into account that the electrons inside
the dot are described by a thermal distribution
and the effects of microwave field should be
considered in the frame of the Kubo formalism
for the density matrix (see e.g. \cite{gilles}).
For analysis it is convenient to use 
creation, annihilation operators defined by usual relations
${\hat x} = \sqrt{\frac{\hbar}{2 m \omega_x }} 
({\hat a} + {\hat a}^+)$, ${\hat y} = \sqrt{\frac{\hbar}{2 m \omega_y }}
({\hat b} + {\hat b}^+)$ and 
${\hat p}_x = -i \sqrt{\frac{m \hbar \omega_x}{2}} ({\hat a} - {\hat a}^+)$, ${\hat p}_y = 
- i \sqrt{\frac{m \hbar \omega_y}{2}} ({\hat b} - {\hat b}^+)$.
The  unperturbed Hamiltonian in absence of microwave field takes the form
${\hat H}_0 = \hbar \omega_x ( {\hat a}^+ {\hat a} + 1/2) + 
\hbar \omega_y ( {\hat b}^+ {\hat b} + 1/2)$.
Then the orbital momentum is
\begin{eqnarray}
\nonumber
{\hat L} = \frac{i \hbar}{2 \sqrt{\omega_x  \omega_y}} 
[  (\omega_x - \omega_y)({\hat a} {\hat b} - {\hat a}^+{\hat b}^+) + \\
\label{eq7}
(\omega_x + \omega_y)({\hat a} {\hat b}^+ -{\hat a}^+ {\hat b}) ] .
\end{eqnarray}
The only non zero matrix elements of ${\hat L}$ are 
\begin{eqnarray}
\nonumber
&& <n_x,n_y| {\hat L} |n_x+\delta_x,n_y+\delta_y> = \\
\nonumber
&& \quad \frac{i \hbar}{2 \sqrt{\omega_x \omega_y}} (\delta_x \omega_x - \delta_y \omega_y) \times \\
\label{eq8}
&& \quad [(n_x+(1+\delta_x)/2)(n_y+(1+\delta_y)/2)]^{1/2} ,
\end{eqnarray}
where  $\delta_{x,y}=\pm 1$ and $n_{x,y}$ are oscillator level numbers.
The perturbation induced by a microwave field 
${\hat V} (t)= (- f_x {\hat x} - f_y {\hat y})\cos \omega t = 
 {\hat V} \cos \omega t$
should be also expressed via operators 
${\hat a}, {\hat a}^+, {\hat b}, {\hat b}^+$.

In the Kubo formalism the evolution of the density matrix ${\hat \rho}(t)$
is described by the equation
\begin{equation}
\label{eq9}
      \begin{array}{ll}
i \hbar \frac{\partial {\hat \rho}}{\partial t} = 
[ {\hat H}_0 + {\hat V}(t), {\hat \rho} ] - i \hbar \gamma ({\hat \rho} - {\hat \rho}_0)
      \end{array}
\end{equation} 
where ${\hat \rho}_0$ is the equilibrium density matrix: \\
${\hat \rho}_0 = \sum_{n_x, n_y \ge 0} \rho_{n_x, n_y} |n_x, n_y><n_x, n_y|$.
Using perturbation theory ${\hat \rho}(t)$ can be expanded in powers of 
the external potential amplitude ${\hat V}(t)$ : 
${\hat \rho}(t) = {\hat \rho}_0 + {\hat \rho}_1(t) + {\hat \rho}_2(t) + ...$.
In the first order we have
\begin{eqnarray}
\nonumber
&&<\alpha | {\hat \rho}_1(t) | \beta> = \frac{(\rho(\epsilon_\beta) -\rho(\epsilon_\alpha)) <\alpha|V|\beta> }{\epsilon_\beta - \epsilon_\alpha - \hbar \omega + i \gamma \hbar} e^{i \omega t}/2 + \\
\label{eq10}
&&\frac{(\rho(\epsilon_\beta) -\rho(\epsilon_\alpha)) <\alpha|V|\beta> }{\epsilon_\beta - \epsilon_\alpha + \hbar \omega + i \gamma \hbar} e^{-i \omega t}/2 \quad .
\end{eqnarray}
For the harmonic potential we obtain
\begin{eqnarray}
\nonumber
&&<n_x+\delta_x, n_y| {\hat \rho}_1(t) | n_x, n_y> = \\
\label{eq11}
&&(\rho_{n_x,n_y}-\rho_{n_x+\delta_x,n_y})
f_x \sqrt{\frac{n_x+(1+\delta_x)/2}{8 \hbar m \omega_x }}  \times \\
\nonumber
&&\left(\frac{e^{i \omega t}}{-\omega_x \delta_x - \omega + i \gamma} + 
\frac{e^{-i \omega t}}{-\omega_x \delta_x + \omega + i \gamma} \right) \; ,
\end{eqnarray}

\begin{eqnarray}
\nonumber
&&<n_x, n_y+\delta_y| {\hat \rho}_1(t) | n_x, n_y> = \\ 
\label{eq12}
&&(\rho_{n_x,n_y}-\rho_{n_x,n_y+\delta_y}) 
f_y \sqrt{\frac{n_y+(1+\delta_y)/2}{8 \hbar m \omega_y }}  \times \\
\nonumber
&&\left(\frac{e^{i \omega t}}{-\omega_y \delta_y - \omega + i \gamma} + 
\frac{e^{-i \omega t}}{-\omega_y \delta_y + \omega + i \gamma} \right) \quad .
\end{eqnarray}
The time averaged second order correction to the density matrix is given by
\begin{equation}
\label{eq13}
 <\alpha | <{\hat \rho}_2(t)>_t | \beta>= 
\frac{<\alpha|<[{\hat V}(t), {\hat \rho}_1(t)]>_t|\beta>}
{\epsilon_\beta - \epsilon_\alpha + i \hbar \gamma} 
\end{equation} 
where $[...]$ marks the commutator between two operators.

To compute the average momentum $L$ we need to find the terms
$<n_x+\delta_x, n_y+\delta_y| <{\hat \rho}_2(t)>_t | n_x, n_y>$.
According to (\ref{eq13}) they are expressed via the matrix elements like
\begin{eqnarray}
\nonumber
&&< <n_x+\delta_x, n_y+\delta_y | {\hat V}(t) | n_x+\delta_x, n_y>  \times \\
\label{eq14}
&&<n_x+\delta_x, n_y| {\hat \rho}_1(t) | n_x, n_y> >_t  = \\
\nonumber
&& (\rho_{n_x,n_y}-\rho_{n_x+\delta_x,n_y}) f_x f_y \times \\
\nonumber
&&[(n_y+(1+\delta_y)/2) 
(n_x+(1+\delta_x)/2)/(64 m^2 \omega_x \omega_y)]^{1/2} \times \\
\nonumber
&&\left( \frac{1}{-\omega_x \delta_x - \omega + i \gamma} + \frac{1}{-\omega_x \delta_x + \omega + i \gamma} \right) ,
\end{eqnarray}
and
\begin{eqnarray}
\nonumber
&&< <n_x+\delta_x, n_y+\delta_y | {\hat \rho}_1(t) | n_x, n_y+\delta_y> \times \\
\label{eq15}
&&<n_x, n_y+\delta_y| {\hat V}(t)  | n_x, n_y> >_t  = \\
\nonumber
&& (\rho_{n_x,n_y+\delta_y}-\rho_{n_x+\delta_x,n_y+\delta_y}) f_x f_y \times \\
\nonumber
&&[(n_y+(1+\delta_y)/2) (n_x+(1+\delta_x)/2)/(64m^2 \omega_x \omega_y)]^{1/2} \times \\
\nonumber
&& \left( \frac{1}{-\omega_x \delta_x - \omega + i \gamma} + 
\frac{1}{-\omega_x \delta_x + \omega + i \gamma} \right) \quad .
\end{eqnarray}
Therefore
\begin{eqnarray}
\nonumber
&& < <n_x+\delta_x, n_y+\delta_y | {\hat V}(t) | n_x+\delta_x, n_y> \times \\
\label{eq16}
&& <n_x+\delta_x, n_y| {\hat \rho}_1(t) | n_x, n_y> >_t - \\
\nonumber
&& < <n_x+\delta_x, n_y+\delta_y | {\hat \rho}_1(t) | n_x, n_y + \delta y> \times \\
\nonumber
&& <n_x, n_y + \delta_y| {\hat V}(t)  | n_x, n_y> >_t = \\
\nonumber
&&[(n_y+(1+\delta_y)/2) (n_x+(1+\delta_x)/2)/(64m^2 \omega_x \omega_y)]^{1/2} \times \\
\nonumber
&& f_xf_y \left( \frac{1}{-\omega_x \delta_x - \omega + i \gamma} + \frac{1}{-\omega_x \delta_x + \omega + i \gamma} \right) \times \\
\nonumber
&& (\rho_{n_x,n_y} - \rho_{n_x+\delta_x,n_y} - \rho_{n_x,n_y+\delta_y} + \rho_{n_x+\delta_x,n_y+\delta_y}   )
\end{eqnarray}
and 
\begin{eqnarray}
\nonumber
&&<n_x+\delta_x, n_y+\delta_y| <{\hat \rho}_2(t)>_t | n_x, n_y> = \\
\label{eq17}
&&\frac{f_x f_y}{8 \hbar m \sqrt{\omega_x \omega_y}} g_{n_x, n_y, \delta_x, \delta_y}(\omega) \times \\\nonumber
&&[(n_x+(1+\delta_x)/2)(n_y+(1+\delta_y)/2)]^{1/2} \quad ,
\end{eqnarray}
where
\begin{eqnarray}
\nonumber
&&  g_{n_x, n_y, \delta_x, \delta_y}(\omega) = \\
\label{eq18}
&& [ \frac{1}{-\omega_x \delta_x - \omega + i \gamma} + \frac{1}{-\omega_x \delta_x + \omega + i \gamma} + \\
\nonumber
&&\frac{1}{-\omega_y \delta_y - \omega + i \gamma} + \frac{1}{-\omega_y \delta_y + \omega + i \gamma} ] \times \\
\nonumber
&&\frac{\rho_{n_x,n_y} - \rho_{n_x+\delta_x,n_y} - \rho_{n_x,n_y+\delta_y} +
\rho_{n_x+\delta_x,n_y+\delta_y}}{- \omega_x \delta_x - \omega_y \delta_y + i \gamma} \; .
\end{eqnarray}
Thus
\begin{eqnarray}
\nonumber
&& <n_x, n_y | {\hat L}| n_x+\delta_x, n_y+\delta_y> \times \\
\label{eq19}
&& <n_x+\delta_x, n_y+\delta_y| <{\hat \rho}_2(t)>_t | n_x, n_y> = \\
\nonumber
&& \frac{i f_x f_y}{16 m \omega_x \omega_y}  G_{n_x,n_y,n_x+\delta_x,n_y+\delta_y}(\omega) \; ,
\end{eqnarray}
where
\begin{eqnarray}
\nonumber
&& G_{n_x, n_y, \delta_x, \delta_y}(\omega) =  \\
\label{eq20}
&& [\frac{1}{-\omega_x \delta_x - \omega + i \gamma} + \frac{1}{-\omega_x \delta_x + \omega + i \gamma} + \\
\nonumber
&& \frac{1}{-\omega_y \delta_y - \omega + i \gamma} + \frac{1}{-\omega_y \delta_y + \omega + i \gamma}] \times \\ 
\nonumber
&&
\frac{\delta_x \omega_x - \delta_y \omega_y}{- \omega_x \delta_x - \omega_y \delta_y + i \gamma} \times \\
\nonumber
&& (\rho_{n_x,n_y} - \rho_{n_x+\delta_x,n_y} - \rho_{n_x,n_y+\delta_y} + \rho_{n_x+\delta_x,n_y+\delta_y}) \times \\
\nonumber
&&(n_x+(1+\delta_x)/2)(n_y+(1+\delta_y)/2) \; .
\end{eqnarray}
Therefore, the final result is
\begin{eqnarray}
\nonumber
&&<L> = Tr({\hat L} <{\hat \rho_2(t)}>)  \\
\nonumber
&& = \frac{i f_x f_y}{16 m \omega_x \omega_y} \sum_{n_x, n_y \ge 0, \delta_x, \delta_y} G_{n_x,n_y,n_x+\delta_x,n_y+\delta_y}(\omega)\\
\nonumber
&&= \frac{f_x f_y}{16 m \omega_x \omega_y} I(\omega) \sum_{n_x, n_y \ge 0} \rho_{n_x,n_y} \\
\label{eq21}
&&= \frac{f_x f_y}{16 m \omega_x \omega_y} I(\omega) N \; ,
\end{eqnarray}
where
\begin{eqnarray}
\nonumber
&& I(\omega) = i \sum_{\delta_x, \delta_y} [ \frac{1}{-\omega_x \delta_x - \omega + i \gamma} + \frac{1}{-\omega_x \delta_x + \omega + i \gamma} + \\
\label{eq22}
&& \frac{1}{-\omega_y \delta_y - \omega + i \gamma} + \frac{1}{-\omega_y \delta_y + \omega + i \gamma} ] \times \\
\nonumber 
&& \frac{\delta_x \omega_x - \delta_y \omega_y}{- \omega_x \delta_x - \omega_y \delta_y + i \gamma} \delta_x \delta_y \; .
\end{eqnarray}
Here $N$ is the number of electrons in the quantum dot.
Of course,  $I(\omega)$ is real and can be presented by another equivalent 
expression:
\begin{equation}
\label{eq23}
I(\omega) = \frac{Q(\omega)}{R(\omega)} \; ,
\end{equation}
with
\begin{eqnarray}
\nonumber
&&Q(\omega) = 8 \gamma \omega_x \omega_y (\omega_x^2 - \omega_y^2) (5 \gamma^6+20 \gamma^4 \omega^2 + 9 \gamma^2 \omega^4 +  \\
\label{eq24}
&&2 \omega^6 + (6 \gamma^4 + 15 \gamma^2 \omega^2 - \omega^4) \omega_y^2 + \\
\nonumber
&&(\gamma^2 + 3 \omega^2) \omega_y^4 + \omega_x^4\ (\gamma^2 + 3 \omega^2 + \omega_y^2) + \\
\nonumber
&& \omega_x^2 (6 \gamma^4 + 15\gamma^2 \omega^2 - \omega^4 + (7 \gamma^2 - 8 \omega^2) \omega_y^2 + \omega_y^4))
\end{eqnarray}
and
\begin{eqnarray}
\nonumber
&&R(\omega) = ((\gamma^2 + \omega^2)^2 +  2(\gamma^2 - \omega^2) \omega_x^2 + \omega_x^4) \times \\
\label{eq25}
&&((\gamma^2 + \omega^2)^2 + 2(\gamma^2 - \omega^2)\omega_y^2 + \omega_y^4) \times \\
\nonumber
&&(\omega_x^4 + \omega_y^4 + \gamma^4 + 2[ (\omega_x^2 + \omega_y^2) \gamma^2 - \omega_x^2 \omega_y^2 ])  \; .
\end{eqnarray}
We used the relation 
\begin{eqnarray}
\nonumber
&&(n_x+(1+\delta_x)/2)(n_y+(1+\delta_y)/2)+ \\
\nonumber
&&(n_x+(1-\delta_x)/2)(n_y+(1-\delta_y)/2) - \\ 
\label{eq26}
&&(n_x+(1-\delta_x)/2)(n_y+(1+\delta_y)/2)- \\
\nonumber
&&(n_x+(1+\delta_x)/2)(n_y+(1-\delta_y)/2) = \delta_x \delta_y
\end{eqnarray}
to reduce Eqs.(\ref{eq19}),(\ref{eq20}) to Eqs.(\ref{eq21})-(\ref{eq23}).

It is important to note that the final result (\ref{eq22})
for the average momentum $L$ is independent of unperturbed thermal
distribution $\rho_{n_x,n_y}$. The momentum grows linearly with the
number of electrons in the dot $N$. For $\omega \sim \gamma \sim \omega_x \sim \omega_y$
we have $L \sim f_xf_y N/(m \omega^3)$ in agreement with the phenomenological
result (\ref{eq4}). However, the exact dependence (\ref{eq22})-(\ref{eq25})
obtained here from the Kubo theory 
is different from the phenomenological result (\ref{eq4}) obtained 
originally in \cite{chaplik}. For example,
at $\omega \gg \omega_x \sim \omega_y > \gamma$
our result gives 
$L \sim \gamma N/(m\omega^2(\omega_x^2-\omega_y^2))$
while the phenomenological result (\ref{eq5}) of \cite{chaplik}
gives $L \propto 1/\omega^6$.

\begin{figure}
\vglue 0.4cm
   \centering
   \includegraphics[height=0.33\textwidth,angle=0]{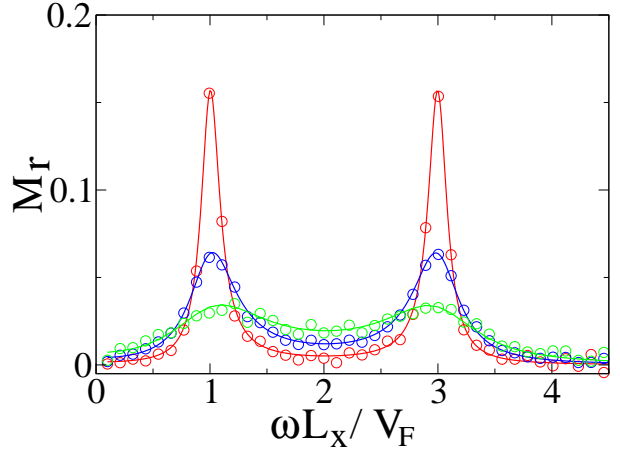}
\vglue 0.5cm
\caption{(color online) Dependence of the rescaled magnetization
$M_r=-2 M/M_0$ on the rescaled microwave frequency
$\omega/\omega_x$ for a quantum dot with a
harmonic potential at $\omega_y/\omega_x=3$
(see definitions of $M_0$ and $M_r$ in Eq~(\ref{eq33})). Here
 $L_x = v_F/\omega_x$ and 
$v_F$ is the velocity Fermi at the Fermi energy $E_F$. 
The rescaled relaxation rate is $\gamma/\omega_x=0.1; 0.25; 0.5$
(red, blue, green curves/circles from top to bottom at
$\omega/\omega_x=1$). Circles show numerical data
obtained from Eq.(\ref{eq30})  by integration of classical dynamics
and Monte Carlo averaging over $10^4$ trajectories
from the Fermi-Dirac equilibrium distribution at zero temperature;
full curves show the theoretical result given by 
Eqs.(\ref{eq21})-(\ref{eq25}).
}
\label{fig1}
\end{figure}

To obtain the expression for $L$  we used above the quantum Kubo theory.
However, the result (\ref{eq22}) has a purely classical form
and therefore it is useful to try to obtain it from the classical kinetic theory.
With this aim let us consider  an arbitrary two dimensional system with 
a Hamiltonian $H = H(q_x, p_x, q_y, p_y,t)$. Then the kinetic equation
for the distribution function $\rho(\mathbf{x})$ \cite{pitaevsky} reads 
\begin{equation}
\label{eq27}
\frac{\partial \rho}{\partial t} + \{H, \rho \} = - \gamma (\rho - \rho_0)
\end{equation}
where $\rho_0$ is the equilibrium thermal distribution,
 $\{H, f\}$ are the Poisson brackets and for simplicity of notations
 $\mathbf{x} = (x, p_x, y, p_y)$.
After a change of variables $(t', \mathbf{x'}) = (t, T_{0,t} \mathbf{x})$ this equation 
is reduced to 
\begin{equation}
\label{eq28}
\frac{\partial \rho(\mathbf{x'},t')}{\partial t'} = 
- \gamma (\rho(\mathbf{x'},t') - \rho_0(T_{t',0}\mathbf{x'}))
\end{equation} 
where $T_{t',t}$ notes the time evolution operator from time $t$ to time $t'$
given by the dynamics of the Hamiltonian $H$. 
This equation can be solved explicitly that leads to the time averaged distribution function : 
\begin{equation}
\label{eq29}
<\rho(\mathbf{x},t)>_t = \gamma \int_{-\infty}^{0} e^{\gamma t'} < \rho_0 (T_{t+t',t} \mathbf{x}) >_t d t'
\end{equation} 
where $T_{t+t',t}$ is the time evolution operator in the phase space.
Then the average of any quantity $Q(\mathbf{x})$ (for example $Q = x p_y - p_x y$) 
can now be expressed as :
\begin{eqnarray}
\nonumber
&&<Q(\mathbf{x})> = \int Q(\mathbf{x}) <\rho(\mathbf{x},t)>_t dx = \\
\nonumber
&& \gamma < \int_{-\infty}^{0} e^{\gamma t'} (\int Q(\mathbf{x}) \rho_0 (T_{t+t',t} \mathbf{x}) dx) dt' >_t = \\
\nonumber
&& \gamma \int dx \rho_0 (\mathbf{x})  \int_{-\infty}^{0} dt' e^{\gamma t'} <Q(T_{t,t+t'} \mathbf{x})>_t = \\
\label{eq30}
&& \gamma \int dx \rho_0 (\mathbf{x})  \int_{0}^{\infty} dt' e^{-\gamma t'} <Q(T_{t+t',t} \mathbf{x})>_t \quad .
\end{eqnarray}
Here it is used that the transformation $\mathbf{x'} = T_{t+t',t} \mathbf{x}$
is area-preserving in the phase space.
 For the case of two oscillators with frequencies $\omega_x, \omega_y$ the time evolution
can be find explicitly so that for dynamics in $x$ we have
\begin{eqnarray}
\nonumber
&&\left(
\begin{array}{l}
x(t+t') \\
\frac{p_x(t+t')}{\omega_x}
\end{array}
\right) 
=
\left(
\begin{array}{ll}
\cos \omega_x t'    &   \sin \omega_x t'\\
-\sin \omega_x t'    &   \cos \omega_x t'
\end{array}
\right) \times \\
\label{eq31}
&&\left(
\begin{array}{l}
x - \frac{f_x/m}{\omega_x^2 - \omega^2} \cos \omega t \\
\frac{p_x}{\omega_x} + \frac{f_x \omega }{(\omega_x^2 - \omega^2) \omega_x} \sin \omega t
\end{array}
\right) + \\
\nonumber
&&\left(
\begin{array}{l}
\frac{f_x/m}{\omega_x^2 - \omega^2} \cos \omega (t+t') \\
- \frac{\omega f_x}{(\omega_x^2 - \omega^2) \omega_x} \sin \omega (t+t')
\end{array}
\right)
\end{eqnarray}
with a similar equation for $y,p_y$. After averaging over $t$ we obtain
\begin{eqnarray}
\nonumber
&&<x(t+t') p_y(t+t')>_t = \frac{\omega}{2} \frac{f_x}{(\omega_x^2 - \omega^2)} 
\frac{f_y/m}{(\omega_y^2 - \omega^2)} \times \\
\nonumber
&&[ - \frac{\omega_y}{\omega} \sin (\omega_y t') \cos(\omega_x t') + \sin \omega t'  \cos(\omega_x t') + \\
\label{eq32}
&& \frac{\omega}{\omega_x} \sin (\omega_x t')\cos (\omega_y t')   -  \frac{\omega}{\omega_x} \sin (\omega_x t') \cos \omega t']
\end{eqnarray}
with a similar expression for $<y(t+t') p_x(t+t')>_t$. After substitution of (\ref{eq32})
in (\ref{eq30}) the integration over $t'$ gives exactly the expression (\ref{eq22})
with $I(\omega)$ given by Eqs.(\ref{eq23})-(\ref{eq25}). The integration can be done
analytically or with help of Mathematica package. This result shows that the average
momentum $L$ can be exactly obtained from the classical formula (\ref{eq30}).

A comparison of results of numerical simulations
of classical dynamics with Monte Carlo averaging over large
number of trajectories from an equilibrium distribution
is shown in Fig\ref{fig1}. The numerical data clearly confirm
the validity of the theoretical expressions given by 
Eqs.(\ref{eq22})-(\ref{eq25}). 

It is interesting to note that
if instead of Eq.(\ref{eq9}) for the density matrix
one would assume an adiabatic switching of microwave field
with a rate $\gamma$ then the average induced momentum $L$
would be given by the classical relation (\ref{eq4})
for a classical oscillator. 
Indeed, such a procedure simply induces an imaginary 
shift in driving frequency.
Such type of switching had been assumed
in \cite{chaplik} and may be considered to give a qualitatively
correct result even if a rigorous description is given by
Eq.(\ref{eq9}) with the final answer in the form
of Eqs.(\ref{eq21}-\ref{eq25}) being quantitatively different from
 Eq.(\ref{eq9}).

For comparison with the physical values of magnetic moment $M$ in real
quantum dots it is convenient to use rescaled momentum $M_r$.
To do this rescaling we note that  the magnetic moment 
is expressed via the  orbital momentum as
$M = e L / (2 mc)$ (in SGS units). Due to the relation (\ref{eq21})
it is convenient to choose a unit of orbital momentum
induced by a microwave field for one electron as 
$L_0=  m v_F L_x (f_x f_y L_x^2)/E_F^2$
where $E_F=m v_F^2/2$ is the Fermi energy in a dot.
Then the unit of magnetic momentum is
$M_0 = e N L_0/mc$ where $N$ is the number of electrons in a dot.
This implies that the physical magnetic moment 
$M$ can be expressed via our rescaled value
according to the relations
\begin{eqnarray}
\nonumber
M= - M_r M_0/2 , \; M_0 = e N L_0/mc , \; \\
\label{eq33}
  M_0 = e N v_F L_x (f_x f_y L_x^2)/(c E_F^2) .
\end{eqnarray}
where the oscillation frequency in $x$ is $\omega_x=V_F/L_x$.
It is important to stress that the total magnetization 
of the dot is proportional to the number of electrons in a
dot with fixed $E_F$.

\section{Dots with an enharmonic potential}

It is very important to extent the methods developed above to 
a generic case of enharmonic potential inside a dot.
With this aim we consider the case of 
2D quartic nonlinear oscillator described by the Hamiltonian
\begin{equation}
\label{eq34}
H=(p_x^2+p_y^2)/2m+(r_x^2 x^4 + r_y^2 y^4+2Kx^2y^2)/4 .
\end{equation}
For $K=0$ we have two decoupled quartic oscillators.
Due to nonlinearity the frequencies of oscillations
scale with energy as $\omega_{x,y} \approx 1.2 (r^2_{x,y} E_F/m^2)^{1/4}$
(see e.g. \cite{chirikov1979}). In this case
$L_x=(4E_F/r_x^2)^{1/4}$, $L_y=(4E_F/r_y^2)^{1/4}$
and we choose the fixed ratio $L_y/L_x=\sqrt{r_x/r_y} = 1/\sqrt{3}$ for our studies.
The rescaled momentum and magnetization are again given by Eq.(\ref{eq33}). 
\begin{figure}
\vglue 0.5cm
\vglue 0.4cm
   \centering
   \includegraphics[height=0.33\textwidth,angle=0]{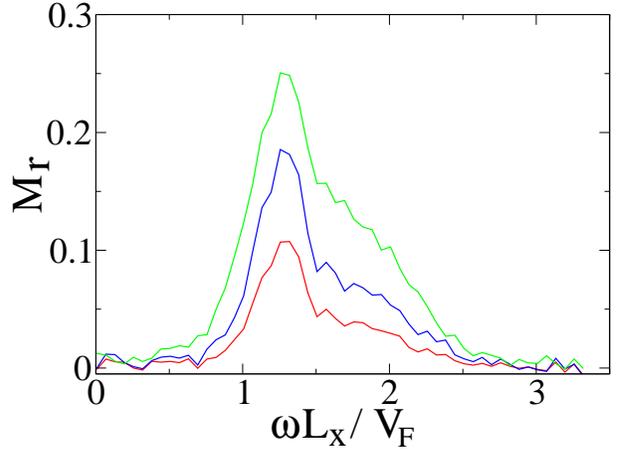}
\caption{(color online) Dependence of rescaled momentum $M_r$
given by Eq.(\ref{eq33})
on rescaled frequency 
$\omega L_x/v_F$ for the case of quartic oscillator
with the Hamiltonian (\ref{eq34})
at $K=0$ and $L_y/L_x=1/\sqrt{3}$ ($r_x/r_y=1/3$), the monochromatic field
has $f_x=f_y$ and $f_x L_x/E_F=0.25 ??$.
Curves correspond to different values of the relaxation rate
$\gamma L_x/v_F$  = 0.03 (red), 0.06 (blue), 0.14 (green)
(from bottom to top at $\omega L_x/v_F =1.2$).
The curves are obtained from Eq.(\ref{eq30}) by 
integration of classical dynamics and Monte Carlo averaging over $10^4$
trajectories from the Fermi-Dirac distribution at zero temperature.
}
\label{fig2}
\end{figure}

The value of averaged orbital momentum is obtained by numerical integration
of hamiltonian equations of motion
in presence of a linear-polarized monochromatic field
(see (\ref{eq1}), (\ref{eq34})). The effects of relaxation
to stationary state with a rate $\gamma$ are taken into account via relation
(\ref{eq30}). The average momentum $<L>$ is obtained by Monte Carlo averaging
over $10^4$ trajectories from
the Fermi-Dirac distribution at
zero temperature. The integration time is about $10^4$
oscillation periods. 

\begin{figure}
\vglue 0.7cm
   \centering
   \includegraphics[height=0.33\textwidth,angle=0]{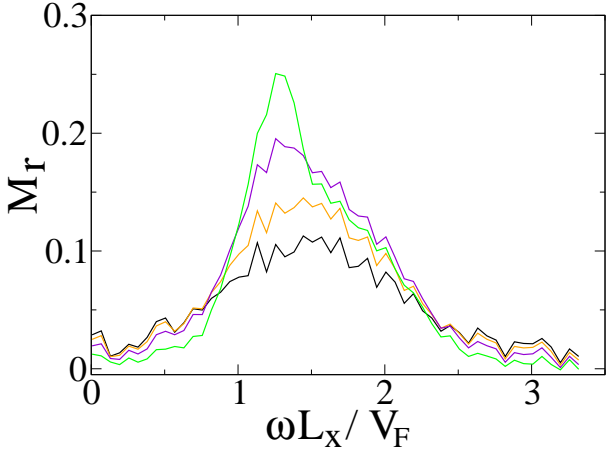}
\caption{(color online) Same as in Fig.2 but for
$\gamma L_x/v_F$ = 0.14 (green), 0.3 (violet),
0.45 (yellow), 0.55 (black)
(from top to bottom at $\omega L_x/v_F =1.2$).
}
\label{fig3}
\end{figure}

The dependence of rescaled magnetization
on microwave frequency $\omega$ for different relaxation rates is shown
in Figs.\ref{fig2},\ref{fig3}. 
In contrast to the case of harmonic potential
the dependence on frequency is characterized by a broad distribution
with a broad peak centered approximately near oscillation frequency
$\omega_x = 1.2(r_x^2 E_F/m^2)^{1/4} \approx  1.2 v_F/L_x$. 
Significant magnetization
is visible essentially only inside the interval
$\omega_x \approx 1.2 v_F/L_x < \omega < \omega_y \approx 2.1 v_F/L_x$.
Data obtained also show that  at small relaxation rates
magnetization drops to zero approximately as
$M_r \propto \gamma$ (see Fig.\ref{fig2}).
Also $M_r$ drops with the increase of $\gamma$ at
large $\gamma$ (see Fig.\ref{fig3}). This behaviour is  qualitatively
similar to the case of harmonic potential (see Eqs.(\ref{eq5}),(\ref{eq6})).

\begin{figure}
\vglue 0.6cm
   \centering
   \includegraphics[height=0.33\textwidth,angle=0]{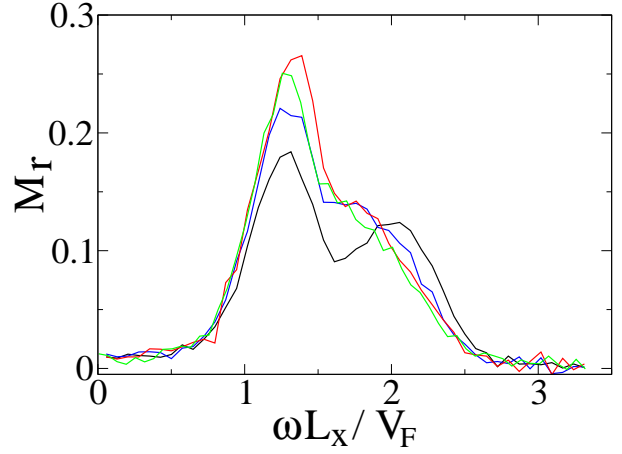}
\caption{(color online) Same as in Fig.2 for
$\gamma L_x/v_F = 0.14 $
and different coupling strength $K$
between $x,y$-modes in (\ref{eq34}):
$K L_x^4/E_F=$ 4 (red), 0 (green), 8 (blue), 16 (black)
(from top to bottom at $\omega L_x/v_F =1.3$).
}
\label{fig4}
\end{figure}

It is important to note that according to data
shown in Fig.\ref{fig4} the coupling between 
$x,y$-modes, which generally leads to a chaotic dynamics \cite{chirikov1979},
does not lead to significant modifications
of the magnetization spectrum. Only at rather large values of $K$,
when the modes are strongly deformed,
the spectrum starts to be modified. 

To show that the magnetization dependence on frequency
found for the quartic oscillator (\ref{eq34})
represents a typical case we also study magnetization in
chaotic billiards described in the next Section.

\section{Magnetization in chaotic billiards}

\begin{figure}
\vglue 1.4cm
   \centering
   \includegraphics[height=0.28\textwidth,angle=0]{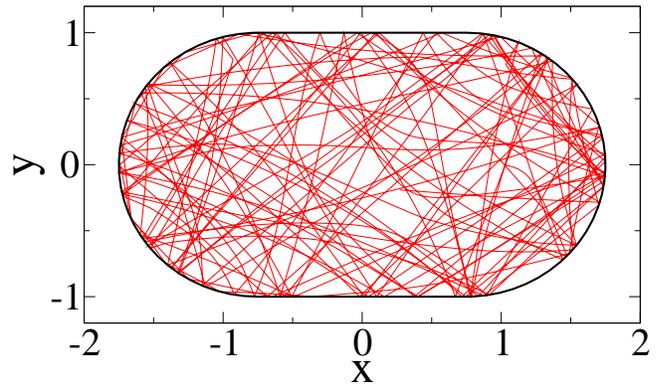}
\caption{(color online) 
Example of a trajectory inside the Bunimovich
stadium dot in presence of microwave
driving. The system parameters are
$L_x/R=3.5, L_y/R=2$, microwave polarization angle $\theta = \pi/4$
($f_x=f_y$), $f R/E_F = 0.28$, $\omega L_x/V_F =1.7$,
$T/E_F=0.1$, $v_F \tau_{rel}/R \approx 2$, $v_F \tau_i/R  \approx 500$.
}
\label{fig5}
\end{figure}

To study microwave induced magnetization in billiards
we choose the Bunimovich stadium billiard \cite{bunimovich}
as a typical example. The semicircle radius is taken to be $R$,
the total size of stadium in $x$ is $L_x$, and in $y$ it is
$L_y=2R$. Usually we use $L_x=3.5R$ (see Fig.\ref{fig5}).
Inside the billiard the particle is affected only by
monochromatic force, the collisions with boundaries
are elastic. 

To take into account
that without monochromatic force the particles relax to
the Fermi-Dirac equilibrium we used the generalized Metropolis
approach developed in \cite{alik}. Namely,
after a time interval $\Delta t$ the kinetic energy of particle
$E$ is changed randomly into the interval $(E - \Delta E, E + \Delta E)$
with a probability given by the Fermi-Dirac distribution
at given temperature $T$. The change takes place only in energy
while the velocity direction remains unchanged (see below).
This procedure imposes a convergence to the
Fermi-Dirac equilibrium with the relaxation
time $\tau_{rel} \approx \Delta t (E_F / \Delta E)^2 $
(see \cite{alik} for detailed description of the algorithm).
In the numerical simulations we usually used
$v_F \Delta t /R \approx 0.05$ and $\Delta E/E_F = 0.15$.
At such parameters a particle propagates on a sufficiently
large distance during the relaxation time: 
$v_F \tau_{rel}/R \approx 2$. 
In absence of $ac-$driving the Metropolis algorithm described above
gives a convergence to the Fermi-Dirac distribution
with a given temperature $T$.
In presence of microwave force the algorithm
gives convergence to a certain
stationary distribution which at small force 
differs slightly from the Fermi-Dirac distribution
(the dependence of deformed curves on energy
is similar to those shown in Fig.\ref{fig2} of Ref.~\cite{alik}).
However, in contrast to the unperturbed case,
the perturbed stationary distribution
has an average nonzero orbital and magnetic
momentum $M$.
The dependence of average momentum $M$ on temperature $T$
is relatively weak if $T \ll E_F$
and therefore,
the majority of data are shown for a typical value
$T/E_F =0.1$ (see more detail below).
To take into account the effect of impurities 
the velocity direction of a particle 
is changed randomly in the interval $(0, 2\pi)$
after a time $\tau_i$. Usually we use
such $\tau_i$ value that the mean free path
is much larger than the size of the billiard
$v_F \tau_i/R  \approx 500$, in this regime
the average momentum is not affected by $\tau_i$
(see below).
The average momentum is usually
computed via one long trajectory
which length is up to $10^7$ times longer than $R$;
computation via 10 shorter trajectories  statistically gives
the same result.
An example of  typical 
trajectory snapshot is shown in Fig.\ref{fig5}.
It clearly shows a chaotic behaviour
(the lines inside the billiard are curved by 
a microwave field).

\begin{figure}
\vglue 0.6cm
   \centering
   \includegraphics[height=0.33\textwidth,angle=0]{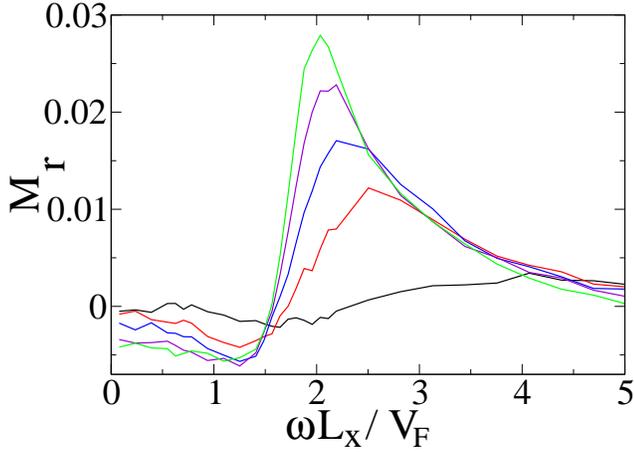}
\caption{(color online) 
Rescaled momentum $M_r$ as a function of 
rescaled microwave frequency $\omega L_x/v_F$
in the Bunimovich billiard. 
All parameters are as in Fig.5,
curves correspond to different values of relaxation
time $\tau_{rel}$ with
$v_F \tau_{rel}/R =$ 
0.5 (green), 0.7 (violet), 1.1 (blue), 2 (red), 8 (black)
(from top to bottom at $\omega L_x/V_F = 2$).
}
\label{fig6}
\end{figure}

\begin{figure}
\vglue 0.7cm
   \centering
   \includegraphics[height=0.33\textwidth,angle=0]{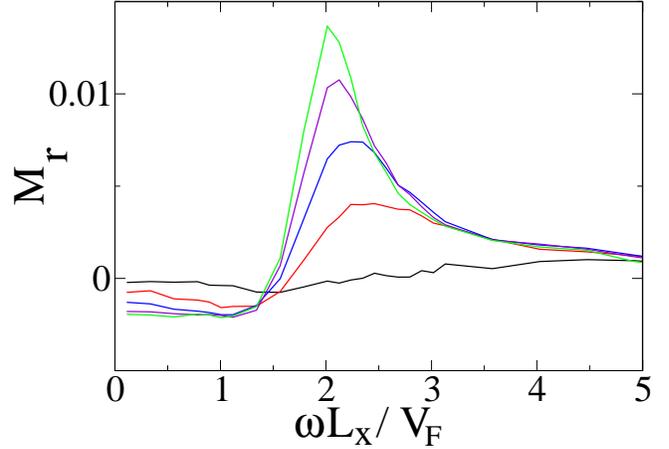}
\caption{(color online) 
Same as in Fig.\ref{fig6} but  the length of stadium 
is increased so that here $L_x/R=5$.
}
\label{fig7}
\end{figure}

The numerical data for dependence of average rescaled momentum $M_r$
(see Eq.~(\ref{eq33})) on rescaled microwave frequency $\omega L_x/v_F$
are shown in Fig.~\ref{fig6} for polarization $\theta=\pi/4$. Qualitatively, 
the dependence is similar to the
case of nonlinear oscillator discussed in previous Section
(see Figs.~\ref{fig2}-\ref{fig4}).
At the same time, there is also a difference in
the behaviour at small
frequency ($\omega L_x/v_F < 1.5$) where the momentum
changes sign. At small relaxation times $\tau_{rel}$
the frequency dependence has a sharp peak near  $\;$
$\omega L_x/v_F \approx 2$. The increase of $\tau_{rel}$
leads to a global decrease of average magnetization,
that is similar to the data of Fig.~\ref{fig2},
also the peak position shifts to a bit higher $\omega$.
Let us also note that according to our numerical
date the rescaled momentum $M_r$ is independent of the strength
of driving force $f$ in the regime when 
$f R/E_F < 0.5$. This is in the agreement with the relation
(\ref{eq33}).

\begin{figure}
\vglue 1.5cm
   \centering
   \includegraphics[height=0.28\textwidth,angle=0]{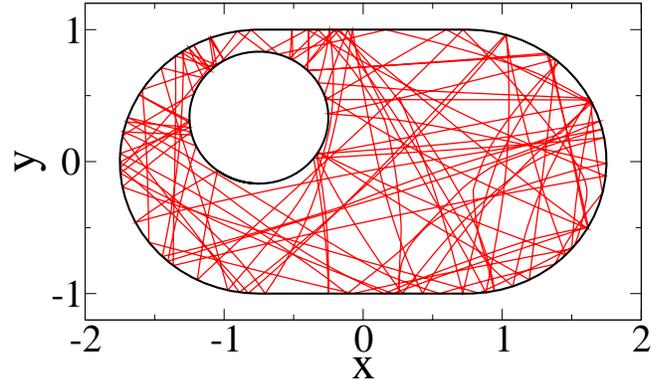}
\caption{(color online) 
Example of a trajectory inside a dot in a form of the Bunimovich
stadium with a circular ``impurity'' inside
the billiard. The billiard boundary is as in Fig.\ref{fig5},
the circular impurity has the
radius $r=R/2$, its center is located at 
$x=-0.75R$, $y=R/3$ counting from the center of the stadium.
The system parameters are as in Fig.\ref{fig5}
except that the polarization angle $\theta=0$.
}
\label{fig8}
\end{figure}

\begin{figure}
\vglue 1.0cm
   \centering
   \includegraphics[height=0.33\textwidth,angle=0]{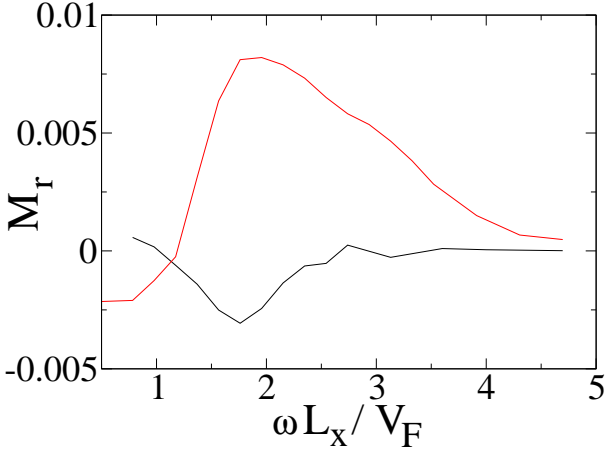}
\caption{(color online) 
Dependence of rescaled momentum
$M_r=-2 M/M_0$ on rescaled frequency
$\omega L_x/v_F$ for the Bunimovich stadium with impurity
shown in Fig.\ref{fig8}.
Here $f R/E_F=0.28$, $\theta=0$ (bottom black curve)
and $f R/E_F=0.2$, $\theta=\pi/4$ (top red/gray curve);
also in this Figure we use 
definition $M_0 = e N v_F L_x (f^2 L_x^2)/(c E_F^2)$ 
which is more suitable for polarization with $\theta=0$.
Other parameters are 
$T/E_F=0.1$, $v_F \tau_{rel}/R \approx 2$, $v_F \tau_i/R  \approx 500$
(cf. with corresponding case in Fig.~\ref{fig6}).
}
\label{fig9}
\end{figure}

The position of peak is determined by the frequency of oscillations
along long $x$-axis of the billiard. Indeed,
an increase of this size of billiard
from $L_x=3.5R$ (Fig.~\ref{fig6}) to $L_x=5 R$ (Fig.~\ref{fig7})
keeps the shape of resonance curves practically
unchanged. At the same time the rescaled magnetization drops approximately
by a factor 2. This means that there is no significant
increase of $M$ with increase of $L_x$
($M_0$ grows by a factor 2.9). 
From a physical view point, it is rather clear since
in the regime with $L_x \gg L_y$ further increase of
$L_x$ cannot lead to increase of magnetization.
This means that in Eq.(\ref{eq33}) the value of $M_0$ gives
a correct estimate of real magnetization assuming that
$L_x \sim L_y$,

The Bunimovich billiard has varies symmetries,
namely $x \rightarrow -x$, $y \rightarrow -y$.
It is interesting to study the magnetization properties
when all of them are absent. With this aim we
introduced an elastic disk scatterer inside the billiard
as it is shown in Fig.~\ref{fig8}.
The dependence of rescaled magnetization $M_r$
on rescaled frequency $\omega L_x/v_F$
is shown for this ``impurity'' billiard
in Fig.~\ref{fig9} for 
two polarization angles $\theta = \pi/4$ and
$\theta=0$. For $\theta = \pi/4$
the behaviour in Fig.~\ref{fig9} is rather similar to 
the case of billiard without impurity
(see Fig.~\ref{fig6}), taking into account
that $M_0$ values differ by a factor two
that gives smaller $M_r$ values in Fig.~\ref{fig9}.
In addition the peak at
$\omega L_x/v_F = 2$ is more broad that
can be attributed 
to contribution of orbits with a shorter periods
colliding with the impurity.
The case of polarization with $\theta=0$
is rather different. Indeed, here the average
magnetization exists even if in the billiard
case $M_r=0$ due to symmetry.
Also the sign of the momentum (magnetization)
is different comparing to the case
of $\theta = \pi/4$ polarization.
For the impurity billiard
we find that at $\theta=0$
the absolute value of magnetization decreases with the 
increase of relaxation time
$\tau_{rel}$ in a way similar to one should
in Figs.~\ref{fig6},\ref{fig7}.

\begin{figure}
\vglue 1.5cm
   \centering
   \includegraphics[height=0.33\textwidth,angle=0]{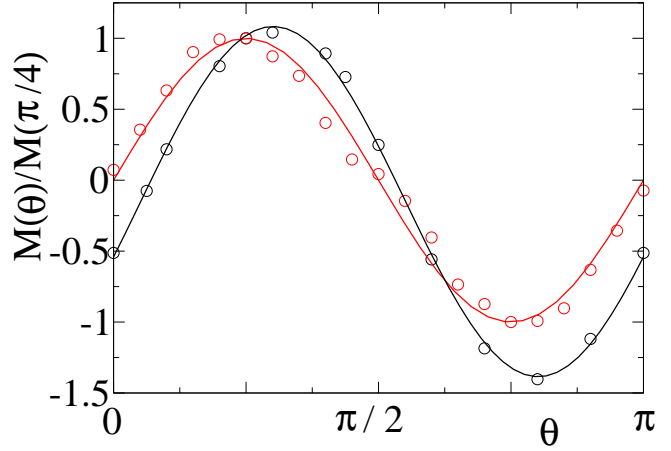}
\caption{(color online) 
Dependence of momentum $M$ on polarization
angle $\theta$ ($M$ is rescaled to its value
at $\theta=\pi/4$)
for the Bunimovich stadium (red/gray)
and the stadium with circular impurity
(black). Numerical data are shown
by open circles obtained 
for $\omega L_x/v_F = 0.78 $ (red/gray)
and $\omega L_x/v_F =1.56$ (black)
at $f R/E_F=0.28$, 
$T/E_F=0.1$, $v_F \tau_{rel}/R \approx 2$, $v_F \tau_i/R  \approx 500$.
Smooth red/gray curve 
shows the theoretical dependence
$\sin 2 \theta$; black curve shows a numerical
fit $M(\theta)/M(\pi/4) = -0.15 + 1.23 \sin(2\theta - 0.32)$.
}
\label{fig10}
\end{figure}

The polarization dependence of magnetization is
shown in more detail in Fig.~\ref{fig10}
for the Bunimovich stadium (Fig.~\ref{fig5}) 
and the impurity billiard (Fig.~\ref{fig8}).
In the first case we have $M(\theta) \propto \sin 2\theta$
as in the case of oscillator so that the magnetization
averaged over all polarization angles is equal to zero.
In contrast to that in the second case
when all symmetries are destroyed
the averaging over all polarization angles gives
nonzero magnetization of the dot. In addition to that
internal impurity gives a phase shift in the
polarization dependence.
The phase shift is due to absence of any symmetry.
In such a case we have a more general dependence
$M \propto (f_xf_y + a f_x^2 +b f_y^2)$,
where $a,b$ are some constants.

\begin{figure}
\vglue 0.7cm
   \centering
   \includegraphics[height=0.28\textwidth,angle=0]{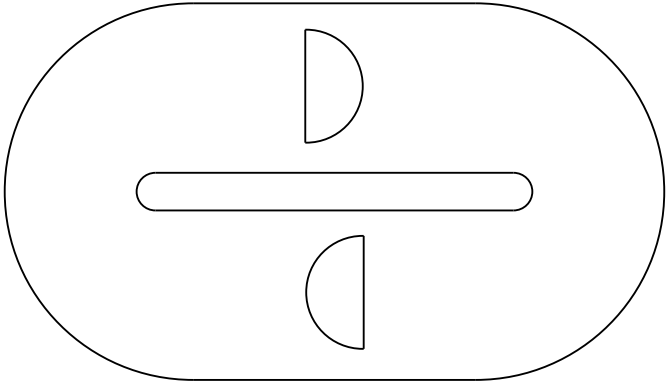}
\caption{Example of a billiard with ratchet effect.
Collisions with the stadium and internal boundaries
are elastic.
}
\label{fig11}
\end{figure}

In addition, we also checked that if
at $\theta=0$ the disk impurity inside the billiard
is replaced by a semidisk of the same radius
then the dependence on the parameters remains essentially
the same, as well as the sign of magnetization
(the disk is divided by a vertical line onto two semidisks
and left semidisk is removed). It is interesting to 
note that a negative sign of $M_r$
means anti-clockwise rotation.
This direction of current rotation can be 
also understood in a link with ratchet
flow on the semidisk Galton board 
studied in \cite{ratchet1,ratchet2,alik}.
The link with the ratchet effect
becomes especially clear if to consider the 
semidisk ratchet billiard
shown in Fig.~\ref{fig11}.
Here, due to the ratchet
effect discussed in \cite{ratchet1,ratchet2,alik} 
for polarization $\theta=0$
electrons should move to the left
in the upper half of the billiard  
and to the right in the bottom half,
thus creating negative magnetization $M_r$.
We think that if only upper semidisk is left
still the direction of rotation at 
$\theta=0$ microwave polarization
is due to the particle flow
directed from right to left in the upper part
of the billiard
that corresponds to negative 
sign of magnetization $M_r$ in Fig.~\ref{fig9}.

Finally, let us make few notes about the
dependence of magnetization $M_r$
on the impurity scattering time $\tau_i$ and 
temperature $T$. For example, for the impurity billiard
of Fig.~\ref{fig8} at $\theta=0$ we find that
the variation of $M_r$ is about $10\%$ 
when $v_F \tau_i/R$ is decreased from $500$ down to $50$,
and then $M_r$ drops approximately linearly with
$v_F \tau_i/R$ (e.g. 
$M_r(v_F \tau_i/R=10)/M_r(v_F \tau_i/R=50)=0.47$,
$M_r(v_F \tau_i/R=5)/M_r(v_F \tau_i/R=50)=0.2$,
$M_r(v_F \tau_i/R=2.5)/M_r(v_F \tau_i/R=50)=0.08$.
The decrease of magnetization with
decrease of $\tau_i$ is rather natural
since the ballistic propagation of a particle
between boundaries disappears as soon as the scattering mean free path
$v_F \tau_i$ becomes smaller than the billiard size $L_x$.
Similar effect has been seen also for the ratchet flow on 
the semidisk Galton board (see Fig.~\ref{fig9} in Ref.\cite{alik}). 
As far as for dependence on temperature
our data show that it was relatively weak
in the regime $T/E_F \ll 1$. For example,
for the set of parameters of Fig.~\ref{fig9} at $\theta=0$
the value of $M_r$ is increased only
by $14\%$ when $T$ is decreased by a factor 4
(from $T/E_F=0.1$ to $T/E_F=0.025$)
and it is decrease by $40\%$ when
$T$ increased from $T/E_F=0.1$ to $T/E_F=0.2$.
This dependence on $T$ shows that
the magnetization of the dot
remains finite even at zero temperature.
This behaviour also has close similarity
with the temperature dependence
for the ratchet flow discussed in
\cite{alik,entin}. Such ballistic dots may
find possible applications for detection
of high frequency microwave radiation 
at room temperatures. Indeed, the ratchet
effect in asymmetric nanostructures,
which has certain links with the magnetization discussed here,
has been observed at 50GHz at room temperature
\cite{asong}.

\section{Discussion}
According to the obtained numerical results (see e.g Fig.\ref{fig2})
and Eq.(\ref{eq33}) the rescaled magnetization of a dot can
be as large as $M_r/2 \sim 0.1$ when the relaxation time
$\tau_{rel}$ is comparable with a
typical time of electron oscillations in a dot.
Thus, the magnetization of a dot 
is $M = 0.1 M_0$. According to Eq.(\ref{eq33}), at fixed electron density
the magnetization  
is proportional to the number of electrons $N$ inside the dot.
Due to this the magnetization induced by a microwave field
can be much larger than the magnetization induced
by persistent currents discussed in
\cite{levy1990,imry,gilles,kravtsov1993,kravtsov2005}.
For example, for 2DES in AlGaAs/GaAs the effective electron mass
$m=0.067m_e$ and at electron density
$n_e \approx 2 \times 10^{11} cm^{-2}$ we have
$E_F \approx 100 K$, $v_F/c \approx 1.4 \times 10^{-3}$.
Hence,  according to Eq.~(\ref{eq33})
for a microwave field of $f/e = 1 V/cm$ acting on an electron
in a dot of size $L_x = 1 \mu m$
we obtain $N \approx 2 \times 10^3$, $f R/E_F \approx 0.01$ and 
$M_0/\mu_B \approx 
2 m_e v_F L_x N (f L_x/E_F)^2/\hbar \approx
5 \times 10^3 N (f L_x/E_F)^2 \approx 10^3$,
where $\mu_B=e \hbar/(2m_e c)$ is the Bohr magneton.
Nowadays technology allows to produce samples with very high mobility
so that the mean free path can have values as large as few tens
of microns at $4K$. At $n_e = const$ we have 
the scaling $M_0 \propto L_x^5 f^2$ and
for an increased dot size $L_x=10 \mu m$ and field $f/e= 3 V/cm$
the magnetization of {\it one} dot is 
$M \approx 0.1 M_0 \approx 10^8 \mu_B$
being comparable with the total magnetization
of $10^7$ rings in \cite{levy1990}.
Therefore, this one dot magnetization induced by a microwave field
can be observed with nowadays experimental possibilities
\cite{bouchiat1,bouchiat2,heitman}.
We note that the magnetization is only weakly dependent on temperature
but the mean free pass at given temperature should be larger
than the dot size. 

This ballistic magnetization should also exist
at very high frequency driving, e.g. THz or optical frequency,
which is much larger than the oscillation frequency.
In this regime the magnetization drops with the  driving
frequency $M \propto 1/\omega^2$ (see Eqs.~(\ref{eq21})-(\ref{eq25}))
but this drop may be compensated by using strong driving fields.

From the theoretical view point many questions remain
open and further studies are needed to answer them.
Thus, an analytical theory is needed to compute
the magnetization in dots with enharmonic potential
or billiards. On a first glance, as a first
approximation one could take analytical formulas for 
harmonic dot Eqs.~(\ref{eq21})-(\ref{eq25})
and average this result over frequencies variation
in an enharmonic dot. However, in the limit of
small relaxation rate $\gamma$ (or large $\tau_{rel}$)
such an averaging gives finite magnetization independent of $\gamma$.
Indeed, in analogy with the Landau damping \cite{landau,pitaevsky}
such integration gives effective dissipation
independent of initial $\gamma \rightarrow 0$.
Appearance of such magnetization independent of $\gamma$
would be also in a qualitative agreement
with the results obtained for ratchet transport
on the semidisk Galton board \cite{ratchet2,alik}
and in the asymmetric scatterer model
studied in \cite{entin}. Indeed, according to these studies
the velocity of ratchet is independent of the
relaxation rate in energy (relaxation over momentum direction
is reached due to dynamical chaos). 
In fact this indicates certain similarities with the Landau
damping where the final relaxation rate
is independent of the initial one. Also, the results obtained
in \cite{alik,entin} show that the ratchet velocity
can be obtained as a result of scattering on one
asymmetric scatterer. 
Therefore, it is rather tentative to use the semidisk billiard
of Fig.~\ref{fig11}
and to say that the magnetization in it appears as the result of
ratchet flow: for polarization $\theta=0$
the ratchet flow goes on the left in the upper part of the billiard
and on the right in the bottom part. The velocity of such
ratchet flow $v_f$ in this 2DES  is given by the 
relations found in \cite{alik,entin}, namely
$v_f/v_F \sim (fR/E_F)^2$. This flow gives a magnetization of 
billiard dot
induced by a microwave field which is of the same order
as $M_0$  in the relation (\ref{eq33}).
However, in this approach the rescaled momentum $M_r$
is simply some constant
independent of the relaxation rate in energy,
as it is the case for the ratchet transport on infinite 
semidisk lattice.
This result
is in the contradiction with our numerical data
for magnetization dependence on the relaxation
rate in energy which gives approximately
$M_r \propto \gamma L_x/v_F \sim L_x/(v_F \tau_{rel})$ 
(see Figs.\ref{fig2},\ref{fig6},\ref{fig7}).
A possible origin of this difference
can be attributed to the fact that
the ratchet flow is considered on an infinite lattice
while the magnetization takes place in a confined system
and the relaxation properties are different in these two cases.

Also we should note that here we used approximation of noninteracting 
electrons and neglected all collective effects.
In principle, it is well known that microwave radiation can
excite plasmons in 2DES (see e.g, \cite{weiss}). These
excitations can be viewed as some oscillatory modes
with different frequencies $\omega_x, \omega_y$
and therefore in analogy with Eqs.~(\ref{eq21})-(\ref{eq25})
it is natural to expect that a liner-polarized 
radiation can also create rotating plasmons
with finite magnetization induced by this rotation.  
In addition, the effects of screening should be also
taken into account.
All these notes  show that further theoretical studies are needed 
for a better understanding of radiation induced 
magnetization in 2DES dots.

We thank Kvon Ze Don for useful discussions and
for pointing to us Ref.\cite{chaplik} at the final stage of this work.
This work was supported in part by the ANR PNANO project MICONANO.

\end{document}